  	\documentclass[twocolumns]{aa}
	\usepackage{graphicx}
 	\usepackage{natbib}
 	
 	\bibpunct{(}{)}{;}{a}{}{,}

\begin{document}

	\title{Interpretation of \textit{HINODE} SOT/SP asymmetric
		Stokes profiles observed in quiet Sun network and internetwork}
	\titlerunning{MISMA interpretation of \textit{HINODE} SOT/SP data}
		
	\author{B.~Viticchi\'{e}\inst{1,2} \and
		J.~S\'anchez Almeida\inst{3} \and
		D.~Del~Moro\inst{2} \and
		F.~Berrilli\inst{2}}
	\authorrunning{Viticchi\'{e} et al.}
	
	\institute{ESA/ESTEC RSSD, Keplerlaan 1, 2200 AG Noordwijk, Netherlands
		\email{bartolomeo.viticchie@esa.int} \and
		Dipartimento di Fisica, Universit\`a degli Studi di Roma
		``Tor Vergata'', Via della Ricerca Scientifica 1, I-00133 Roma, 
		Italy 	
		\email{delmoro@roma2.infn.it, berrilli@roma2.infn.it}
		\and
		Instituto de Astrof\'isica de Canarias E-38205 La Laguna,
		Tenerife, Spain
		\email{jos@iac.es}}

	\begin{abstract}
	{}
	{Stokes profiles emerging from the magnetized solar photosphere 
	and observed by SOT/SP aboard the \textit{HINODE} satellite
	present a variety of complex shapes. These are 
	indicative of unresolved magnetic structures
	that have been overlooked in the inversion analyses performed
	so far. Here we present the first interpretation of the Stokes profile asymmetries 
	measured in the \ion{Fe}{i}~$630$~nm lines by SOT/SP, in both 
         quiet Sun internetwork (IN) and network regions.}
	{The inversion        
    is carried out under the hypothesis of 
    MIcro-Structured Magnetized Atmosphere       
    (MISMA), where the unresolved structure is assumed to be
    optically thin.     
    We analyze a $29.52'' \times 31.70''$ subfield carefully
	selected to be representative of the properties of
	a $302''\times162''$ quiet Sun field-of-view
	at disk center.}
	{
	The inversion 
	code is able to reproduce the 
	observed asymmetries in a very satisfactory way, including the 
	$35$~\% of inverted profiles presenting large asymmetries.
         The inversion code interprets $25$~\% of inverted profiles 
	as emerging from pixels in which both
	positive and negative polarities coexist. These pixels are located
	either in frontiers between opposite polarity
	patches or in very quiet regions. kG field strengths are found 
	at the base of the photosphere
	in
	both network and IN; in the case of the latter, both kG fields and hG fields are admixed.
	When considering the magnetic properties at the 
        mid photosphere most kG fields are gone,
	and the statistics is dominated by hG fields.
         According to the magnetic filling factors derived from the inversion,
    we constrain the magnetic field
	of only $4.5$~\% of the analyzed photosphere
	(and this percentage reduces to $1.3$~\% when 
	referred to all pixels, including those with low polarization 
	not analyzed).
	The rest of the plasma is consistent with the presence of weak fields not 
	contributing to the detected polarization signals. 	
	The average flux densities derived in the full subfield
	and in IN regions are higher than the ones
	derived from the same dataset by Milne-Eddington inversion.}
	 {
	The existence of large asymmetries in \textit{HINODE} SOT/SP
	polarization profiles is uncovered. These are not negligible in quiet
	Sun data. The MISMA inversion code reproduces them in a satisfactory way,
	and provides a statistical description of the magnetized IN and network
	which partly differs and complements the results obtained so far.
	From this it follows the importance of having a complete interpretation
	of the line profile shapes.
	}
	\end{abstract}

	\maketitle
	
\section{Introduction}
\label{intro}
	The spectropolarimeter SOT/SP \citep[][]{Lit01,Tsu08}
	aboard the Japanese \textit{HINODE} satellite \citep{Kos07}
	combines 
	0\farcs32
	angular resolution and $10^{-3}$ polarimetric
	sensitivity to perform seeing-free full Stokes measurements of the
	polarized light emerging from the magnetized solar photosphere.
	Since its launch in $2006$, the SOT/SP instrument has allowed
	the solar community to investigate the solar surface magnetism
	under unprecedented stable conditions.

	\textit{HINODE} SOT/SP observations have been largely
	exploited to investigate the quiet Sun magnetism.
	Its global properties have been described by
	\citet{OroS07}, \citet{Lit08}, \citet{AseR09}, and \citet{Jin09}; while detailed analyses of
	its local properties, in relation with the temporal evolution 
	due to the interaction with plasma motions, have
	been carried out by \citet{Cen07}, \citet{OroS08}, \citet{Nag08}, \citet{Shi08},
	\citet{Fis09} and \citet{Zha09}.

	The above studies have been performed
	exploiting inversion techniques for interpretation of the observed Stokes profiles.
	In most of the cases, the hypothesis of  ME
	atmosphere has been adopted to infer the properties of the magnetic
	field vector in \textit{HINODE} resolution elements.
	It assumes that the polarization is produced in an 
	atmosphere where the magnetic field vector is constant.
	This approximation is a good compromise to interpret large
	data sets in reasonable times, as the ones  
	obtained by \textit{HINODE}.
	However, it is important to realize that \textit{HINODE} 
	angular resolution is still too low to 
	perform spectropolarimetric observations of 
	magnetic structures that can be regarded as 
	uniform \citep[e.g.,][]{SanA04b,Ste09}. As we emphasize in \S~\ref{resexam}, 
	asymmetric Stokes profiles in \textit{HINODE} quiet Sun 
	observations are the rule rather than the exception, 
	which implies the presence of unresolved structure 
	\citep[][and references therein]{SanA96}.
	The ME analyses cannot reproduce them, 
	overlooking the expected coexistence of 
	several
	magnetic components in a single resolution element 
	\footnote{ME codes consider the existence of
	a field-free atmosphere portion in the resolution element via
	a stray-light filling factor \citep[see e.g.,][]{SkuLit87,OroS07}.}.

	\citet{SanA96} put forward arguments
	for describing the atmosphere where the polarization is
	formed as a MIcro-Structured Magnetic Atmosphere (MISMA),
	i.e., an atmosphere in which magnetic fields
	have structure smaller than the photon mean-free path
	at the solar photosphere ($\la100$~km).	
	It can be regarded as a limiting case of structures of all
	sizes \citep[e.g.][]{lan94,car07}. The MISMA hypothesis
	simplifies the radiative transfer, yet it provides realistic
	asymmetric Stokes profiles. Under this 
	hypothesis, the polarization from a single pixel
	in spectropolarimetric observations is equivalent to that
	produced by the average atmosphere. If one 
	considers several components with diverse physical
	properties (i.e., thermodynamics, plasma motions, and 
	magnetic fields), the resulting spectrum is not a	
	linear combination of Stokes profiles emerging from each 
	magnetic component. Rather, the superposition is non-linear, giving rise 
	to asymmetries with the properties observed in the 
	Sun (e.g., with spectral lines that produce net circular
	polarization).
	\citet{SanALit00} showed how such a complex scenario can be properly
	adapted using a three component model MISMA,
	whose implementation in an inversion code
	\citep{SanA97} allowed them to reproduce the full 
	variety of profile asymmetries emerging from the quiet Sun
	when observed with the instrumentation available at the time.
	Stokes profiles in IN and network observations
	performed with \textit{HINODE} exhibit very important 
	asymmetries which encouraged us to attempt the same MISMA analysis on
	these data.
	
	We present an orderly inversion 
	of Stokes profiles observed by 	
	\textit{HINODE} SOT/SP in IN
	and network regions. It is performed using
	the inversion code in \citet{SanA97},
	that has been recently employed to explain
	the reverse polarity patches found 
	by \textit{HINODE} in sunspot penumbrae \citep{SanAIch09}.
	Our work represents the first analysis of quiet 
	Sun SOT/SP data that is able to reproduce the asymmetries of 
	the Stokes profiles. It allows
	us to obtain information
	contained in Stokes profiles that are hidden 
	to the ME analysis. In particular, we constrain the 
	physical properties of the unresolved magnetic structure 
	creating the asymmetries.

	The paper is structured as follows: in \S~\ref{dssel} 
	we briefly present the dataset and
	the selection of a sample subfield representative of the whole
	field-of-view (FOV). In \S~\ref{hypstra} the
	inversion hypotheses are exposed in
	detail, together with the adopted inversion 
	strategy, and a few inversion examples (\S~\ref{resexam}). 
	The	main results from the analysis are presented
	in \S~\ref{res}. These are then discussed in \S~\ref{disc}.
	An outline of conclusions is given in \S~\ref{conc}.
	
\section{Dataset and subfield selection}
\label{dssel}
	\begin{figure}[!ht]
	\centering
 	\includegraphics[width=5cm]{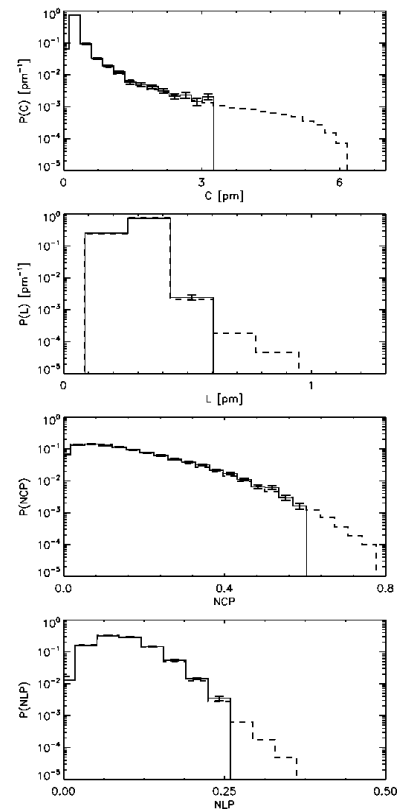}
	\caption{Comparison between the statistical properties of the 
			full $302''\times162''$ FOV (dashed line) and the statistical
			properties of the selected $29.52''\times31.70''$ 
			subfield (solid line). From top to bottom: histogram of
			circular polarization ($C$, first plot), 
			total linear polarization ($L$, second plot),
			net-circular polarization ($NCP$, third plot),
			and net-linear polarization ($NLP$, fourth plot). 
			The error bars are defined as the
			square root of the number of counts in each bin.}
	\label{fig1}
	\end{figure}
	\begin{figure*}[!ht]
	\centering
 	\includegraphics[width=10cm]{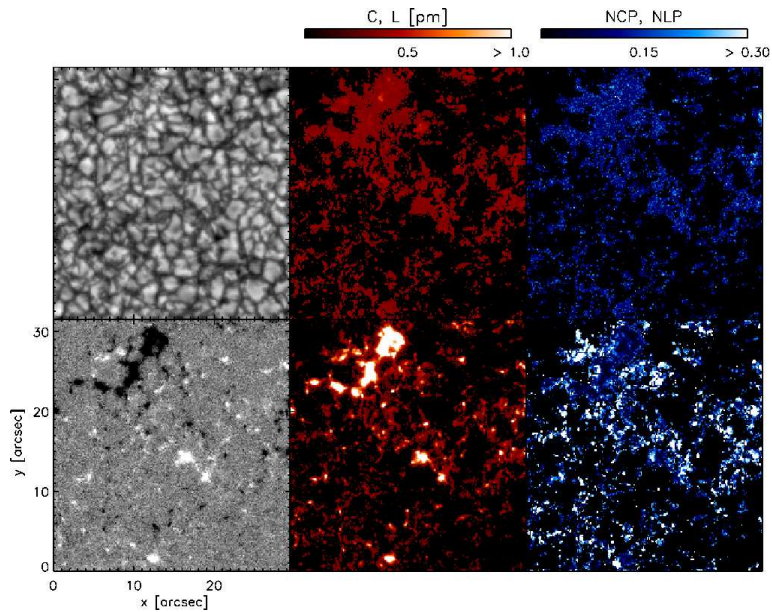}
	\caption{Main observational properties of the representative
			$29.52'' \times 31.70''$ subfield
			selected for in-depth analysis. Continuum intensity (upper-left panel),
			total linear polarization ($L$, upper-central panel),
			net-linear polarization ($NLP$, upper-right panel), COG
			magnetogram saturated at $\pm200$~G (lower-left panel), total circular
			polarization ($C$, lower-central panel), and net-circular polarization
			($NCP$, lower-right panel). The total polarization maps are both saturated at $1$~pm,
			while the net-circular and net-linear polarization maps are saturated at $0.3$.
			Black pixels in $L$, $C$, $NLP$, and $NCP$ maps represent regions 
			with signals below the threshold for inversion.
			The bars on top represent the color palettes adopted for the two
			pairs of images below them}						
	\label{fig2}
	\end{figure*}
	We analyze full Stokes profiles
	of a $302''\times162''$ portion of the 
	solar photosphere observed at disk center on 
	2007 March 10 between 11:37 and 14:34~UT.
    The spectropolarimetric measurements
	have been taken by the SOT/SP instrument
	aboard \textit{HINODE} in the two \ion{Fe}{i} $630$~nm
	lines, with a wavelength sampling of $2.15$~pm~pixel$^{-1}$, and 
	a spatial sampling of $0.1476''$~pixel$^{-1}$ and $0.1585''$~pixel$^{-1}$
	along the east-west and south-north directions, respectively.
	The dataset has been already analyzed by \citet{OroS07},
	\citet{Lit08} and \cite{AseR09} to derive 
	magnetic properties of IN and network regions.
	
	The data reduction and calibration have been performed
	using the \texttt{sp\_prep.pro} routine available in solar soft
	\citep{Ich08}. For a correct absolute wavelength calibration, a correction
	for the gravitational redshift of $613$~m~s$^{-1}$ has been
	taken into account. 
	Using the polarization signals in continuum wavelengths, we estimated	
	a noise level of $\sigma_V\simeq 1.1\times10^{-3}~I_c$
	for Stokes $V$, and of $\sigma_{Q}\simeq \sigma_{U}\simeq 1.2\times10^{-3}~I_c$ 
	for Stokes $Q$ and $U$ ($I_c$ stands for the continuum intensity).
	
	The inversion is carried out on a 
	$29.52''\times31.70''$ subfield (equivalent to $200\times200$~pixels) 
	of the full FOV.
	MISMA inversions are time consuming, and this 
	selection permits the in-depth analysis of a representative FOV 
	in a reasonable timescale.
	The representative subfield has been selected based on two 
	criteria. First, we consider invertible only those Stokes 
	profiles presenting a maximum amplitude in Stokes $V$ or $Q$ or $U$ above 4.5
	times their noise. This check is performed in two different wavelength
	windows centered on the two spectral lines\footnote{Each wavelength window is defined 
	by two wavelengths ($\lambda_A$ and $\lambda_B$)
	centered around the wavelength of the Stokes $I$ minimum ($\lambda_m$).
	The width of the window centered on the \ion{Fe}{i}~630.15~nm line
	is $86$~pm, while for \ion{Fe}{i}~630.25~nm the width is 
	$64$~pm.}.
	Second, the subfield must be a good representation of the whole
	set of invertible data attending to the statistical properties of the total circular polarization
	($C=\int \sqrt{V(\lambda)^2}$d$\lambda/I_c$), total linear polarization
	($L=\int \sqrt{Q(\lambda)^2+U(\lambda)^2}$d$\lambda/I_c$), net-circular
	polarization ($NCP=\vert\Delta V\vert/\int\vert V(\lambda)\vert$d$\lambda$)\footnote{
	$\Delta S=\int_{\lambda_A}^{\lambda_m}\vert S(\lambda)\vert$d$\lambda-
	\int_{\lambda_m}^{\lambda_B}\vert S(\lambda)\vert$d$\lambda$, where $S$ can be $Q,U$, or $V$,
    and $\lambda_A$, $\lambda_B$, and $\lambda_m$ are defined in footnote 
    \footnotemark[2].}, 
	and net-linear polarization ($NLP=(\vert\Delta Q\vert+\vert\Delta U\vert)/
	\int\vert Q(\lambda)\vert+\vert U(\lambda)\vert$d$\lambda$)\footnotemark[6].
	The distributions employed to characterize the statistical
	properties of the full FOV and subfields use 30 bins
	to sample the full domain of each quantity. Bins with
	less than $16$ counts in the case of the subfields, or
	$25$ in the case of the full FOV, are discarded.
	In the subfield selection procedure, all possible $29.52''\times31.70''$ subfields 
	in the $302''\times162''$ FOV are considered. 
	The selected subfield is the one that minimizes the sum of differences 
	between the subfield distributions and the ones defined over the $302''\times162''$ domain. 
	Fig.~\ref{fig1} compares both the $302''\times162''$ statistics and the 
	selected subfield statistics, while Fig.~\ref{fig2} contains
	the selected subfield continuum intensity represented
	together with the center-of-gravity magnetogram 
    \citep[COG; see][]{ReeSem79}, and the $C$, $L$, $NCP$, and $NLP$ maps.

\section{Inversion hypothesis and strategy}
\label{hypstra}
	In order to assign physical properties to each point of
	the portion of photosphere under examination, we work out a model atmosphere
	producing Stokes profiles as close as possible to the observed ones. In 
	the technical literature, this procedure is usually referred to
	as {\em Stokes profile inversion}.
	We perform the inversion under the MISMA hypothesis.
	In a MIcro-Structured Magnetized Atmosphere, magnetic fields
	vary over scales smaller that the photon mean-free-path at the solar photosphere
	\citep{SanA96,SanA97}. Starting from this hypothesis, \citet{SanALit00}
	derived by inversion eleven classes of MISMA models representative of typical profiles observed in
	IN and network regions at disk center.

	All these models are formed by
	three components: one of them is field-free, while the other two are
	magnetized. The field-free component represents the
	non-magnetized plasma in which the magnetic fields are embedded,
	while the two magnetized components allow us to model the 
	coexistence, in the resolution element, of different 
	magnetic fields contributing to the formation of the observed
	polarization profiles.
	It has been shown by \citet[][\S~4.2]{SanA96} that,
	under the MISMA hypothesis, three is the minimum number of components 
	with constant magnetic field required to reproduce the asymmetries of the Stokes
	profiles observed in the network, and three seem to suffice
	to reproduce all quiet Sun profiles \citep[][]{SanALit00}. 
	Such components should be considered as a schematic representation of the
	average properties of the atmosphere 
	\citep[see][\S~5]{SanALit00}.
	Each component of the model is characterized by 
    the variation with height of the thermodynamical properties,
	plasma motions along the magnetic field lines, occupation fraction, and magnetic field 
    strengths. 
    These variations are forced to conserve magnetic flux and mass flows.
    The temperature stratification is imposed to be the same in
    the three optically thin components, since 
	we expect an efficient radiative thermal exchange among them.
	Moreover, the lateral pressure balance links the properties of
	the components at every height; they must 
 	have the same total pressure defined as the
	sum of the gas pressure plus the magnetic pressure. The 
	mechanical balance couples the magnetic field and the thermodynamics in the
	model atmosphere. Magnetic field inclination and azimuth are constant with height.
	Finally, an unpolarized stray-light contribution is considered 
	\citep[for a complete description of the inversion code and 
	hypotheses, see][]{SanA97,SanALit00}. The inversion of the some $400$ wavelengths defining each set
    of Stokes profiles employs only $20$ free parameters, which still
    are twice the number of free parameters of ME inversions \citep[][]{OroS07c}.
	
	The adopted inversion strategy is the following. Stokes $I$ and $V$ are inverted
	in those pixels presenting $\vert V\vert$ amplitude above
	$4.5\times\sigma_{V}$ at least for one wavelength in one of the two \ion{Fe}{i}
	lines. If $\vert Q\vert$ or $\vert U\vert$ amplitude is above $4.5\times\sigma_{Q,U}$,
	at least for one wavelength in one of the two \ion{Fe}{i} lines,
	a full Stokes inversion is also performed.
	According to these criteria about $29$~\% of the
	subfield has been inverted, with 2.3~\% corresponding to full Stokes
	inversions.
	The $4.5\times\sigma_{Q,U,V}$ thresholds have
	been chosen so as to be the same as that of the ME inversion 
	analyses already performed
	on our dataset \citep{OroS07,AseR09}.
	Each profile has been inverted using
	as initial guess all the eleven classes of MISMA models of
	\citet{SanALit00}. Among the eleven inversions, we select the one having the smallest 
	deviation between the observed and the computed profiles.
	Note that when only Stokes~$I$ and $V$ are inverted, we assume the
	magnetic field to be longitudinal for both the magnetic components (i.e., the inclination is either
	zero or $180\degr$). As we argue in Appendix~\ref{appendix2}, this 
	assumption influences the fraction of stray light inferred from the
	inversion, but not the magnetic field strength. 

\subsection{Inversion examples}
\label{resexam}
	In this section we present examples of 
	inversions.
	They have been selected among all the $11600$ inverted profiles, 
	and they are representative of the goodness
	of the fit.
	\begin{figure}[!ht]
	\centering
	\includegraphics[width=5cm]{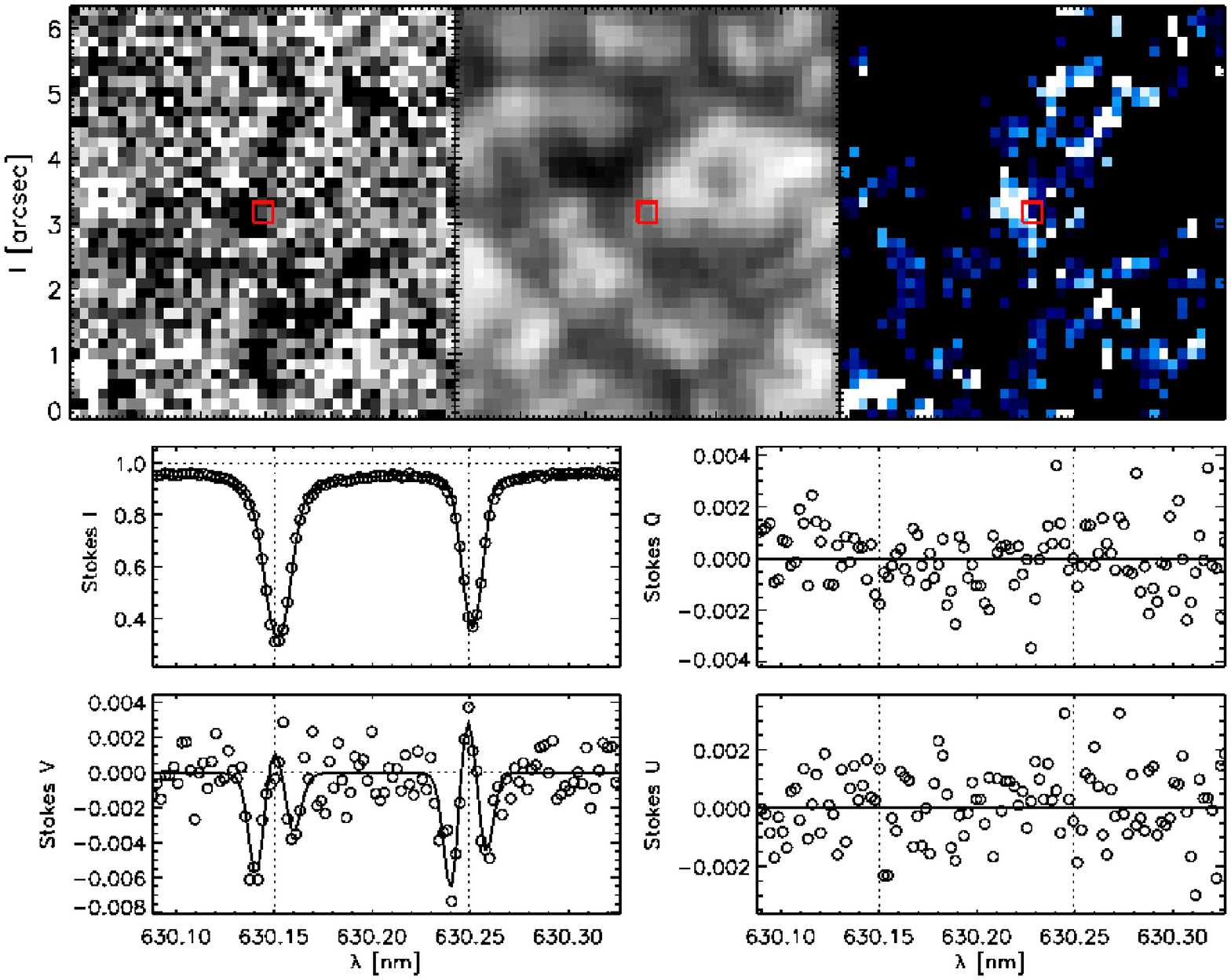}\\
	\includegraphics[width=5cm]{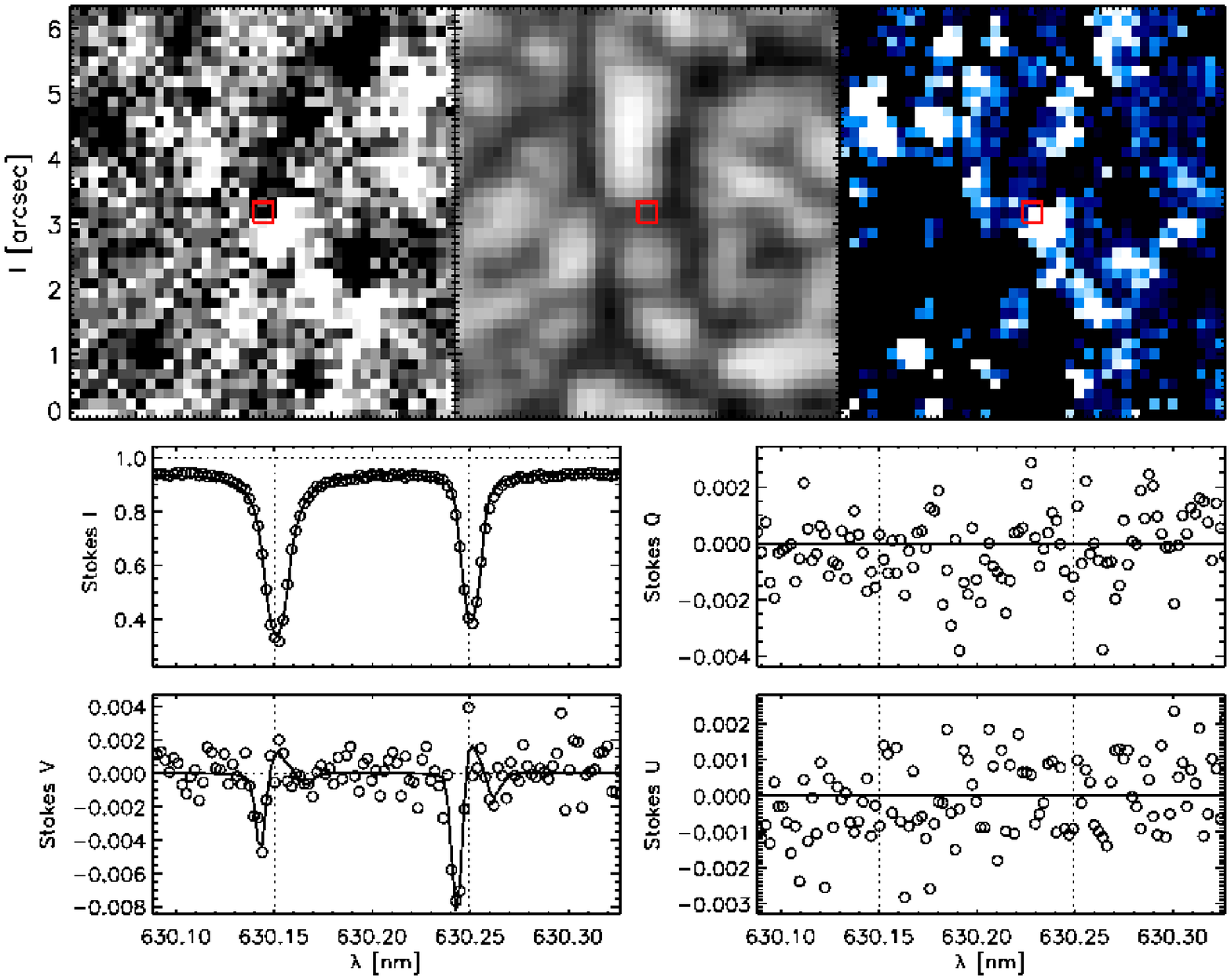}\\
	\includegraphics[width=5cm]{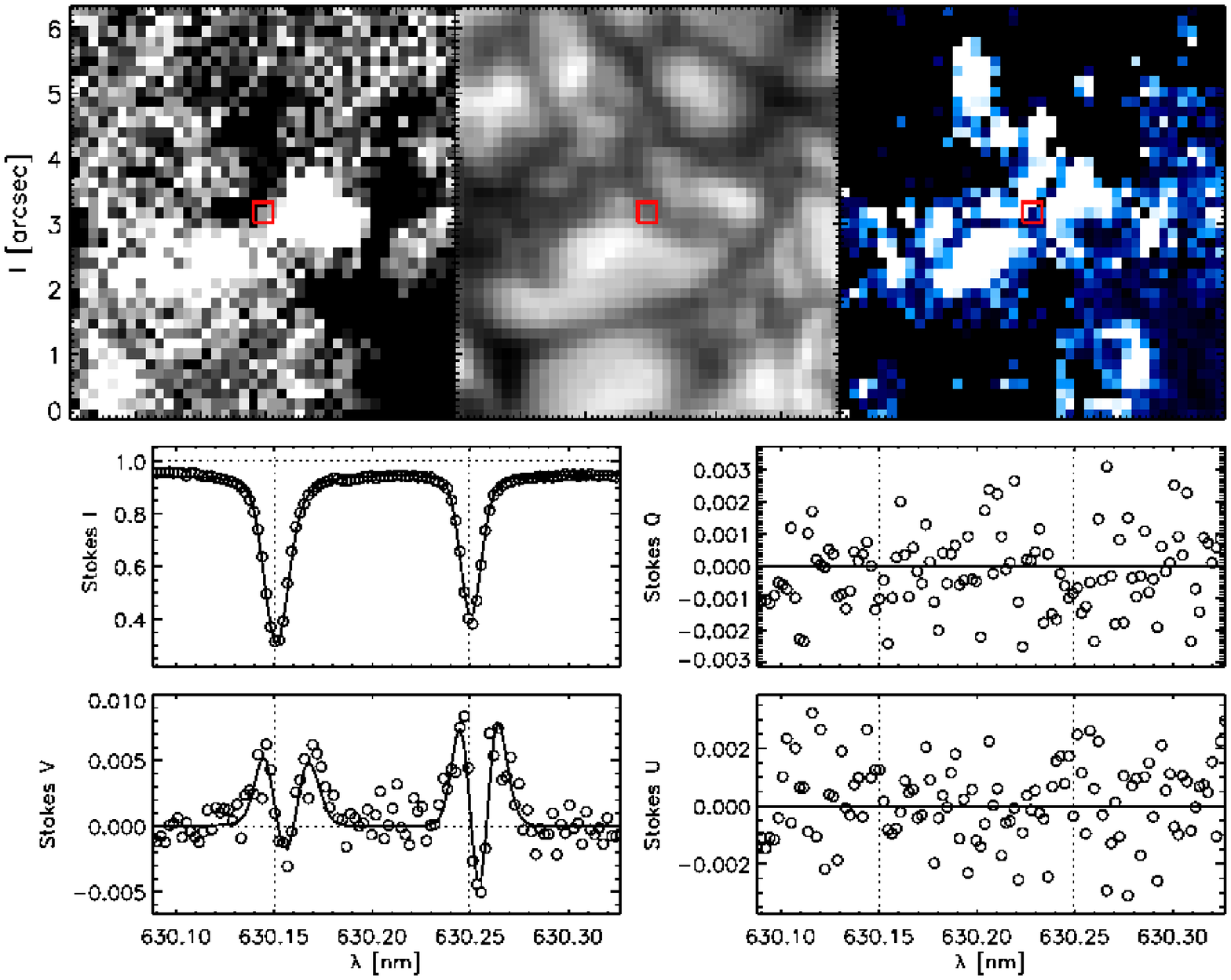}
	\caption{Examples of MISMA inversion of \textit{HINODE} SOT/SP Stokes profiles 
			corresponding to mixed polarities in the resolution element.
			We show profiles in three different pixels representative of
			a weakly polarized IN region
			(upper panels),
			and frontier pixels between regions with opposite polarity 
			in the IN (central panels) and network
			(lower panels). The results are organized
			in two sets of panels; each of these sets contains an upper row
			with images that put the pixel into context, and a lower
			row with the four Stokes profiles (as labeled). The 
			upper row shows $l\times l$ ($l\simeq6''$) maps of the COG magnetogram saturated at $\pm50$~G (left),
			the continuum intensity (central), and the net-circular polarization 
			saturated at $0.3$ (right).
			The red square indicates the position from which the profiles are taken. 
			Lower rows: observed Stokes profiles (symbols) and best-fitting 
			Stokes profiles (solid lines). The vertical dotted lines mark the  
			laboratory wavelengths of the two lines.}
	\label{figpm}
	\end{figure}
	\begin{figure}[!ht]
	\centering
	\includegraphics[width=5cm]{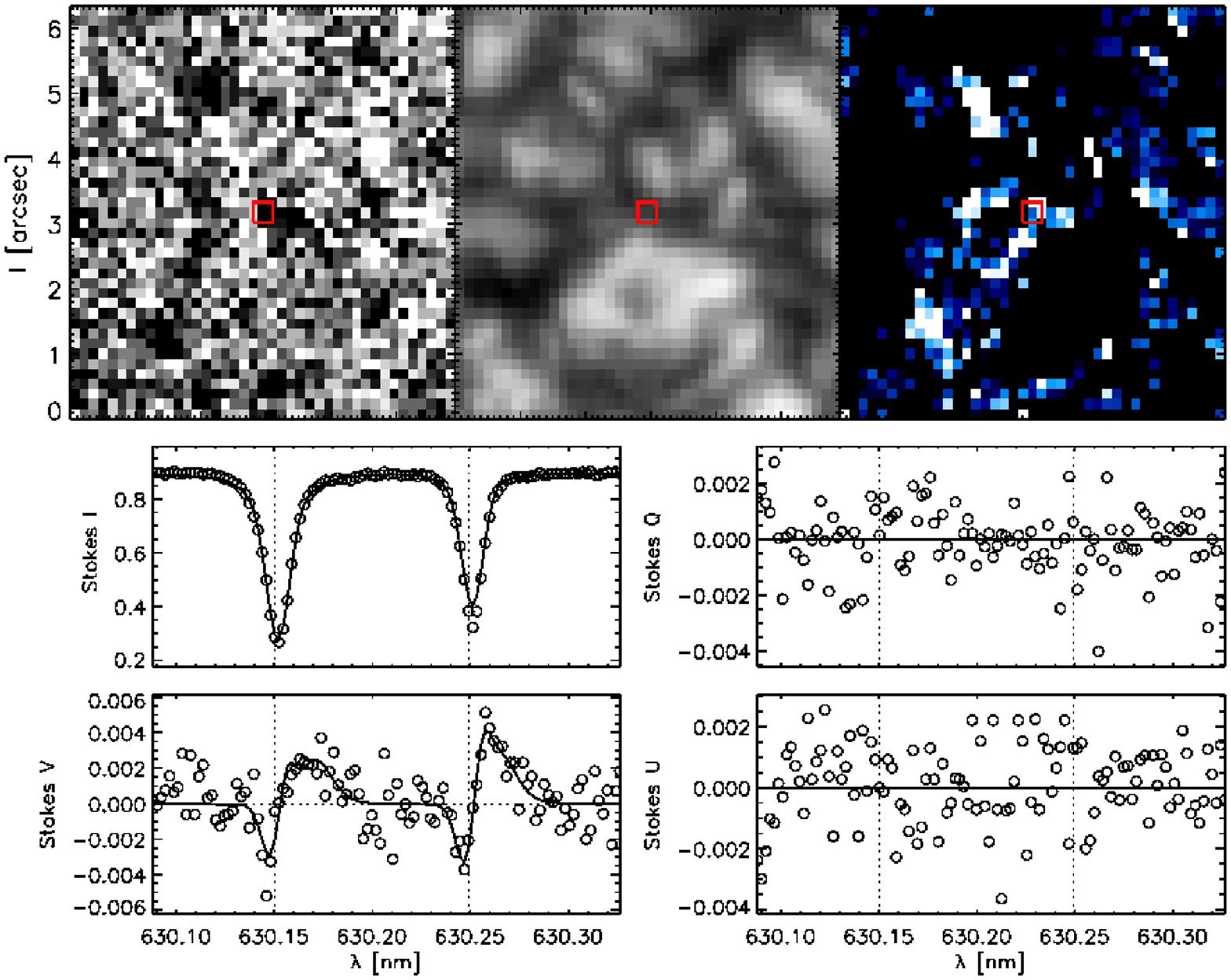}\\
	\includegraphics[width=5cm]{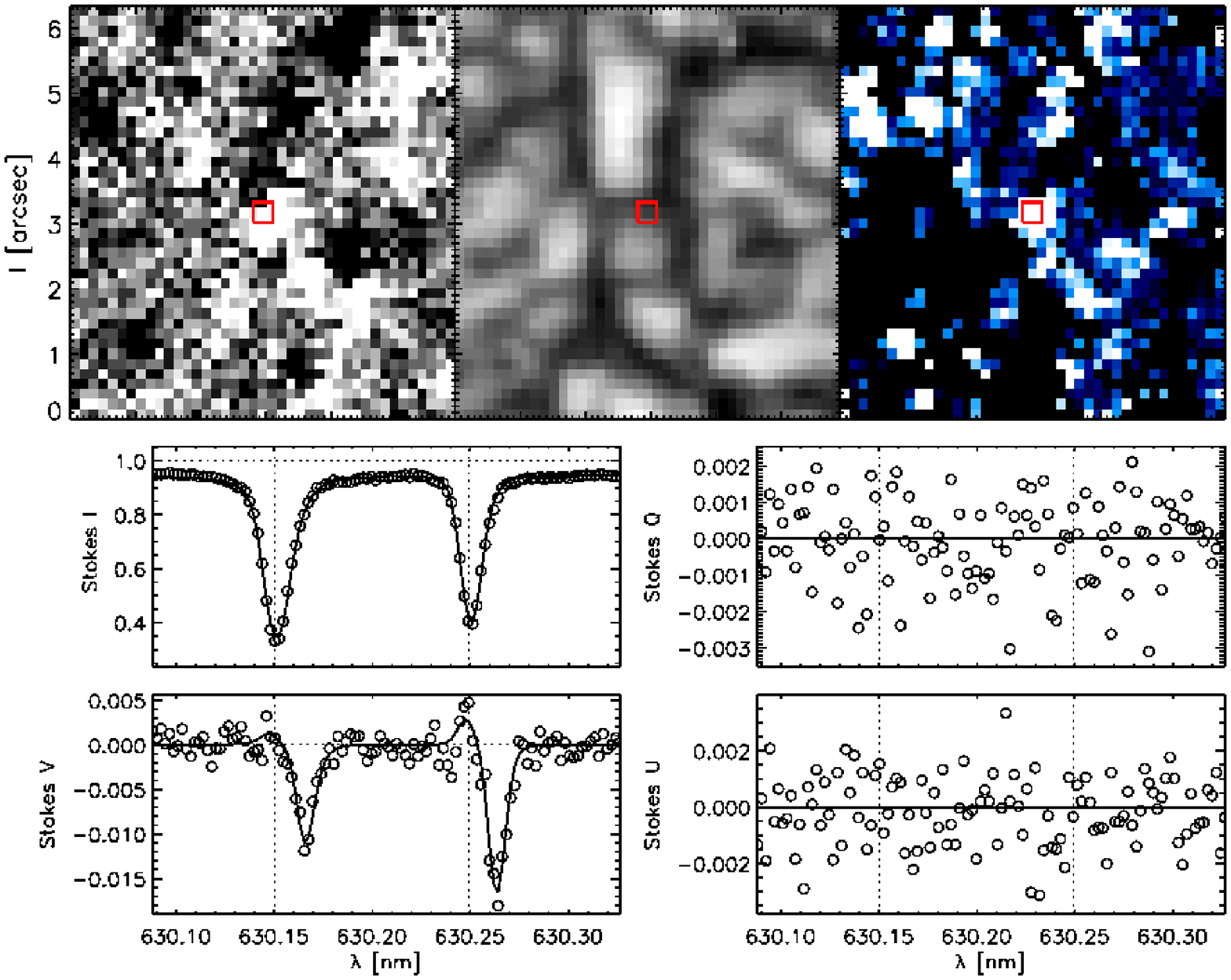}\\
	\includegraphics[width=5cm]{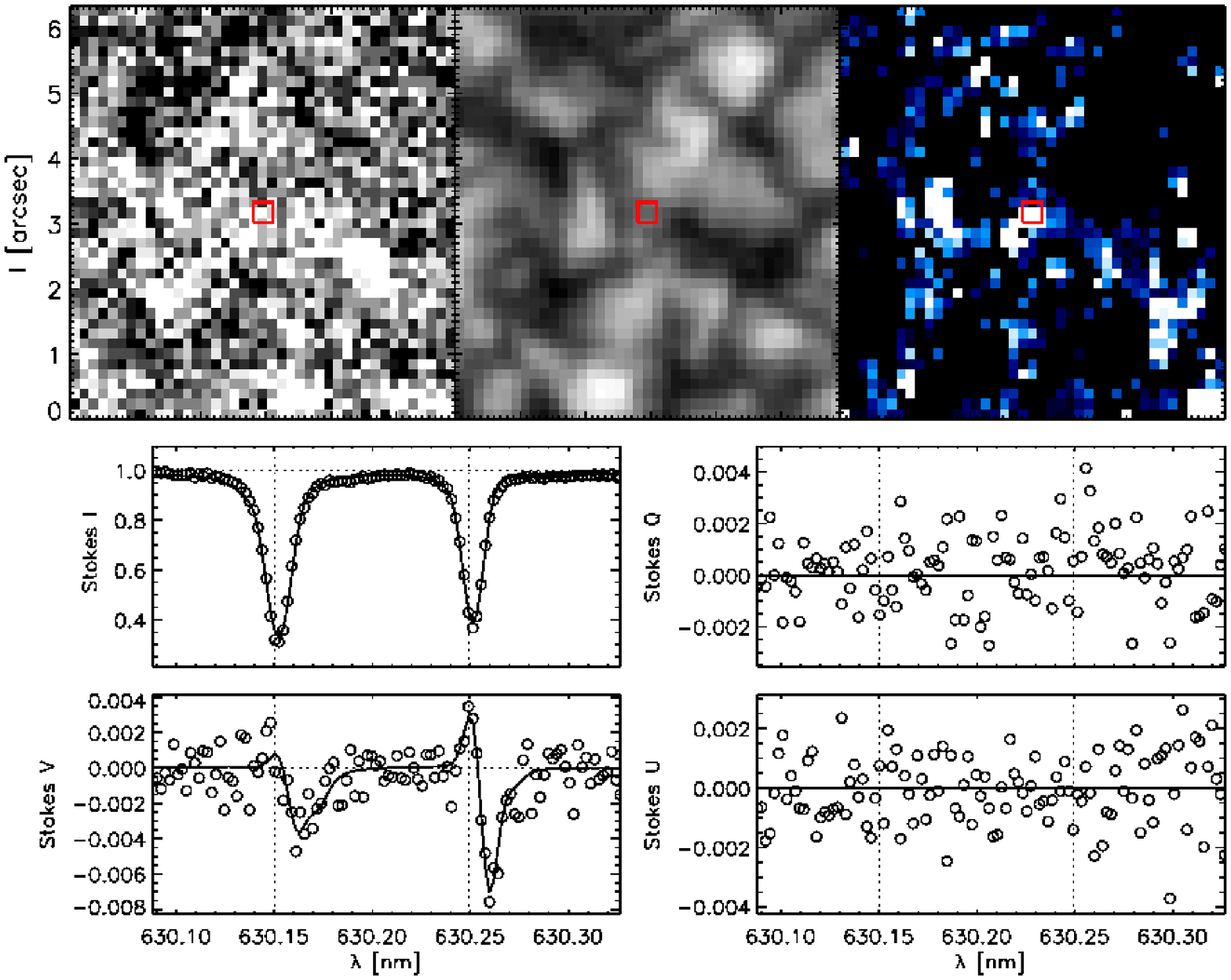}
	\caption{Examples of MISMA inversion of \textit{HINODE} SOT/SP Stokes profiles 
			corresponding to a single polarity in the resolution element.
			For details on the layout of the figure, see the caption of Fig.~\ref{figpm}.}
	\label{figs}
	\end{figure}	
	\begin{figure}[!ht]
	\centering
	\includegraphics[width=5cm]{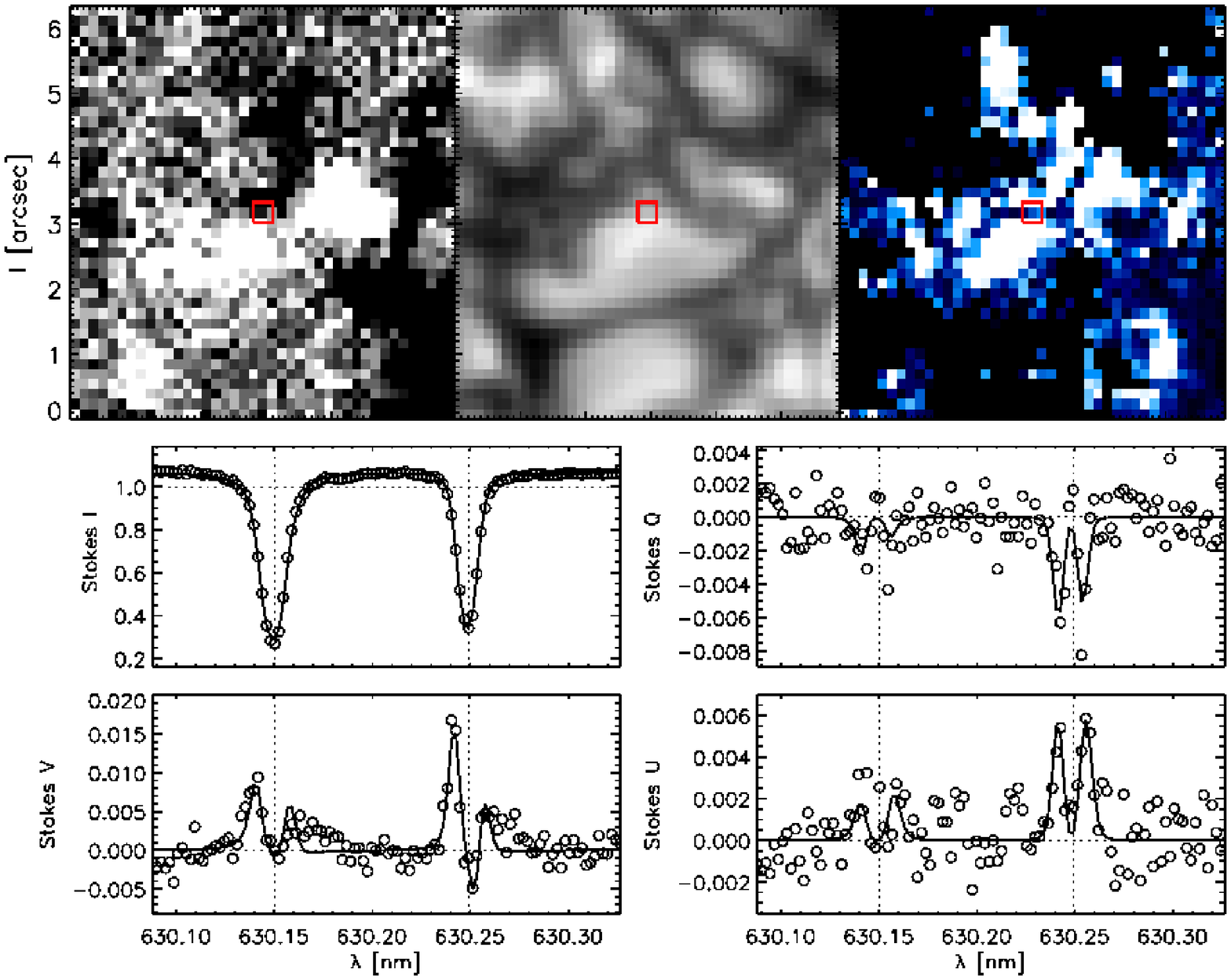}\\
	\includegraphics[width=5cm]{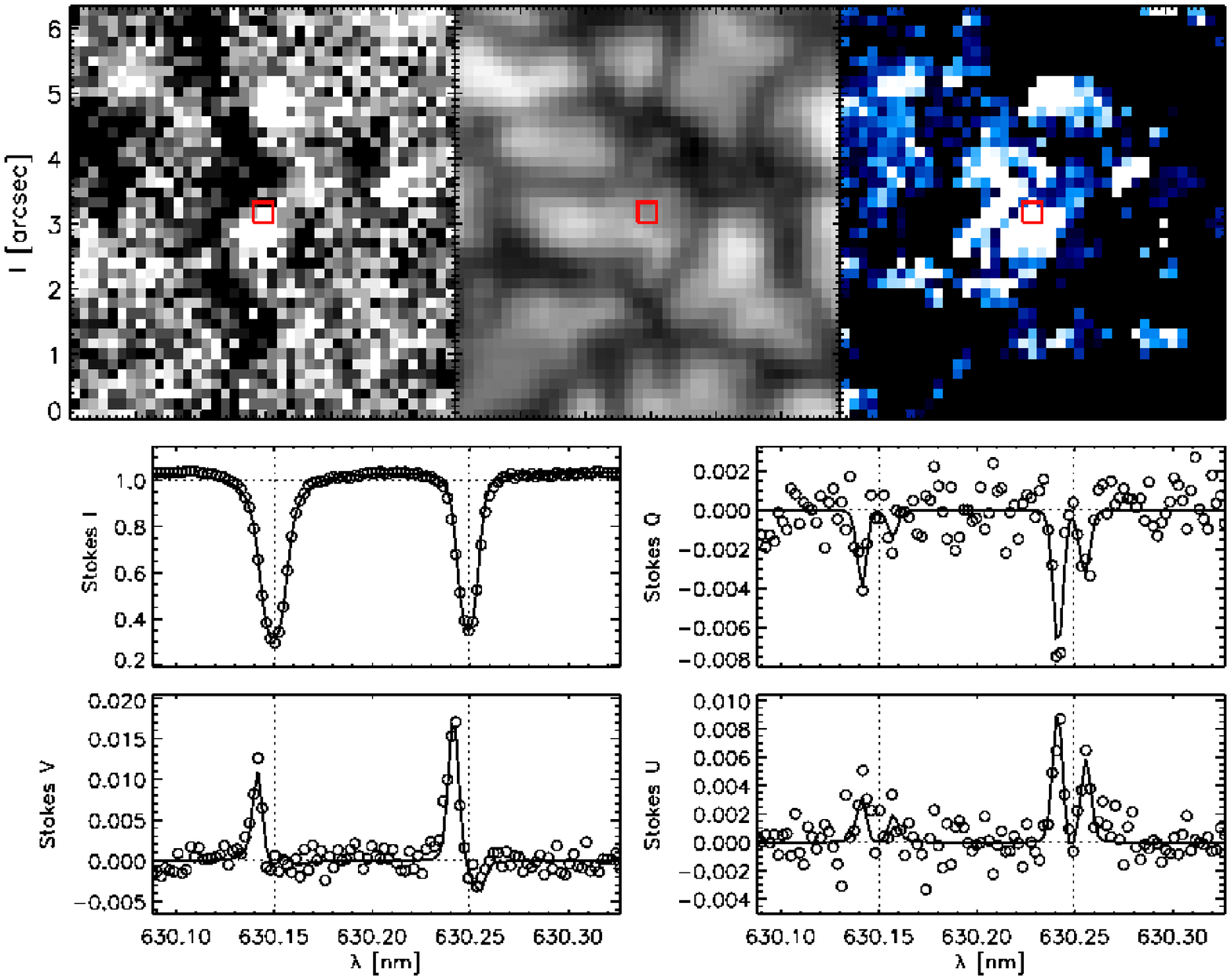}
	\caption{Examples of full Stokes MISMA inversion of \textit{HINODE} SOT/SP Stokes profiles.
			For details on the layout of the figure, see the caption of Fig.~\ref{figpm}.}
	\label{figf}
	\end{figure}

	Fig.~\ref{figpm} shows three examples of 
    pixels where the inversion code retrieves two
	opposite polarity fields coexisting in the resolution element. They
	have been picked in different regions of the
	subfield, i.e., around the negative network
	patch on the top-left corner of Fig.~\ref{fig2}, and in
	IN regions. The profiles show how the two polarities can be either
	easily detectable, when quasi symmetric 
	Stokes~$V$ profiles are measured, or almost 
	hidden, when one of the two lobes of Stokes~$V$
	almost vanishes. Two of the three selected cases
    coincide with a spatially-resolved large-scale
    change of polarity, i.e., the inverted pixels correspond 
    to frontiers between positive and negative 
    polarities in the COG magnetogram. In addition,
    the third example in Fig.~\ref{figpm}  
	shows that mixed polarities
	can be found associated with
	polarized pixels that are not in frontiers between opposite polarities.
	Note that the three cases reported in Fig.~\ref{figpm} show 
	no evidence of linear polarization signals above the
	$4.5\times\sigma_{Q,U}$ threshold. 
	Even below the imposed
	threshold, the linear polarization profiles do not
	present any shape that could be interpreted
	through an inversion procedure with reliability.
	
	Figure~\ref{figs} 
	contains three examples of profiles observed
	far from network regions. They
	have been chosen to be characteristic
	of the typical profiles found in the
	IN as observed by \textit{HINODE} SOT/SP.
	All Stokes~$V$  profiles present asymmetries,
    for which the
	inversion works in a really satisfactory way.
	The central panels of Figs.~\ref{figpm} and \ref{figs} correspond
	to adjacent pixels next to an apparent 
	neutral line in COG magnetogram (i.e., where the magnetic field changes sign; compare
	their position on the magnetogram). They
	have been selected to show how the MISMA inversion discerns between a 
	single polarity model (Fig.~\ref{figs}) and a
	mixed polarity model (Fig.~\ref{figpm}).
	The Stokes $V$ profile in the central panel of 
	Fig.~\ref{figpm} presents a dominant negative blue lobe which indicates
	the presence of negative fields. Beside this, a small negative
	red lobe is also detected by the inversion. The latter denotes 
	the presence of
	positive fields. The MISMA code correctly interprets such a profile
	as emerging from a pixel in which opposite polarities coexist. Such an
	interpretation is consistent with the COG map, showing how
	the pixel from which the profile is taken lies in the frontier
	between positive and negative field patches. The
	central panel of Fig.~\ref{figs} shows the Stokes $V$ profile
	of a pixel next to the previous one in the direction of the 
	positive polarity concentration. It is very different from 
	its neighbor -- the profile in Fig.~\ref{figs} is dominated by
	a negative red lobe and presents a small positive blue lobe.
	The MISMA code interprets such a profile as emerging from
	a pixel in which only positive fields are present. In this 
	case the inversion is also consistent with the COG magnetogram, 
	which shows the pixel lying on a positive polarity patch.
	
	Figure~\ref{figf} presents  two examples of full Stokes inversion, i.e.,
	inversions including $Q$ and $U$. Such examples illustrate the linear
	polarization signals we considered to be invertible.
	The upper panel of Fig.~\ref{figf} shows the inversion of
	a pixel very close to the example in the upper panel of Fig.~\ref{figpm}.
	In this case the linear polarization signal is strong enough
	to be analyzed and the MISMA code succeeds in the inversion.
	It is important to notice that in such a pixel, similarly to
	what found in the upper panel of Fig.~\ref{figpm}, polarization
	profiles are still interpreted by the code as emerging from
	a mixed polarity pixel. In the lower panel of Fig.~\ref{figf}
	a full inversion of an IN pixel is represented. In this case,
	even if the pixel is almost on the frontier between
	opposite polarity regions, a single polarity is measured.
	In the two cases the magnetic field strengths at the base of the 
	photosphere are in the kG regime.
	
	The examples discussed here illustrate 
	not only the goodness of the fits
	but also the soundness of the MISMA interpretation of \textit{HINODE} SOT/SP
	measurements. Such measurements are often characterized
	by important asymmetries in Stokes $V$ profiles; Figs.~\ref{figpm}~and~\ref{figs} show
	three examples of Stokes $V$ profiles whose $NCP\geq0.3$. From the maps in
	Fig.~\ref{fig2}. we notice that such values for the $NCP$ are
	very common in the selected $29.52''\times31.70''$ subfield
	and, by extrapolation, they should be
	very common in the full FOV as well.
	The common presence of
    large asymmetries demands a refined inversion method 
	to interpret quiet Sun SOT/SP profiles, such as the MISMA inversion 
    we are employing.
	Detail on the percentage of asymmetric 
	profiles are reported in \S~\ref{res}.
	
\section{Results}
\label{res}
	\begin{figure*}[!ht]
	\centering
 	\includegraphics[width=10cm]{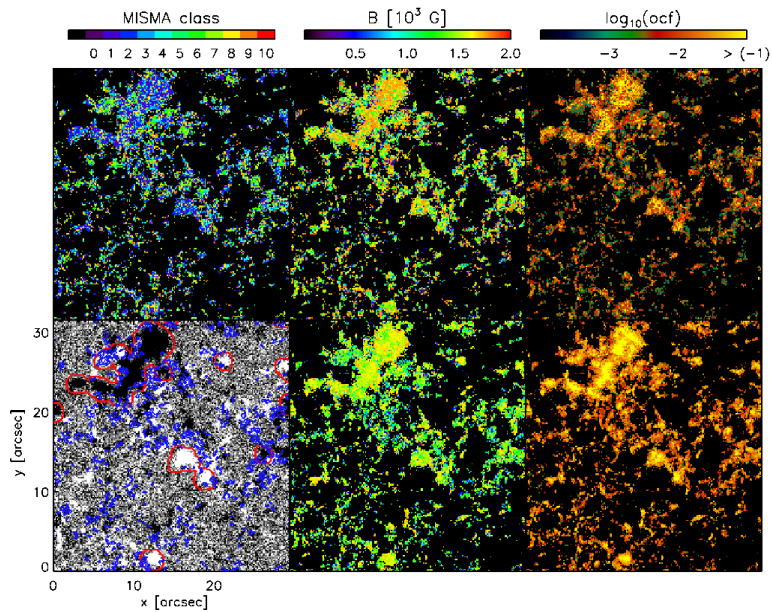}
	\caption{Results of the inversion at the constant reference height, corresponding
			to the base of the photosphere. The different panels contain:
        	classes of initial model MISMAs used in the inversion (upper-left panel),
			field strength for the minor component (upper-central panel),
        	occupation fraction for the minor component (upper-right panel), 
        	COG magnetogram saturated at $\pm50$~G (lower-left panel), field strength
			for the major component (lower-central panel), 
			occupation fraction of the major
			component (lower-right panel). 
			The contours on the COG
			magnetogram represent the network regions as 
			identified by our automatic procedure (red), and
			the pixels with mixed polarities in the resolution element (blue).
			Black pixels in the MISMA class, field strength,
			and occupation fraction maps correspond to regions that are not inverted
			(see Fig.~\ref{fig2}).
			The bars on top represent the color palettes adopted for the MISMA initial models (on the left),
			and for the two pairs of images of field strength (on the center) and occupation fraction (on the right)}.
	\label{fig3}
	\end{figure*}
	The inversion code succeeded in the inversion of $11600$ profiles
	which represent $29\%$ of the selected subfield. 
	The total time needed to perform such an analysis
	is about two days when the inversion analysis is 
	split into eight IDL jobs running in parallel		
	\footnote{The inversion of each pixel
	is independent so the analysis of the invertible domain can be distributed over different CPUs;
	the jobs run on two Xeon E5410 Quad-core processors with 8~GB of shared ram.}.
	Among the $11600$ inverted profiles, $925$ ($\simeq8$~\% of the total
	number) show invertible linear polarization signals.
	These profiles have been found
	mainly in correspondence with patches
	of strong polarization (see the upper panel in Fig.~\ref{figf}).
	Here we present the results obtained from the inversion 
	of Stokes $I$ and $V$ profiles alone.
	Full Stokes inversions using modified MISMA models having inclination and azimuth 
	as free parameters have been performed with success (e.g., Fig.~\ref{figf}), 
	but they are less representative from a statistical
	point of view, and they provide similar results.
	In fact, magnetic field strengths
	from few hG to kG fields are measured in fully inverted
	pixels, with a slightly higher probability for kG fields.
	We focus our analysis on the statistical properties of the
	magnetic field strength and occupation fraction 
	at the photosphere as retrieved by the inversion code
	(the occupation fraction is just the volume filling factor).
	Assuming the model atmospheres to be in lateral pressure
	balance, each value of total pressure defines the same geometrical
	height in all model MISMAs. We decided to take as reference 
	height in the atmosphere the one corresponding to the base of the photosphere
	in typical 1D quiet Sun model atmospheres \citep[e.g.,][]{Mal86}.
	The photospheric lower boundary in such models
	is defined as the height where the 
	continuum optical depth at $500$~nm equals one, which
	corresponds to a pressure of about $1.3\times10^{5}$~dyn~cm$^{-2}$.
	Unless otherwise is mentioned explicitly, all
	the parameters discussed
    hereafter refer to this reference height or,
    equivalently, to this reference total pressure.
	
	Fig.~\ref{fig3} shows six maps summarizing the 
	inversion at the reference height, together with the COG magnetogram, used 
	here to show the context. 
	The red contours on the COG magnetogram 
	separate network and IN regions in the selected subfield.
	Network patches have been identified using an algorithm  which 
	takes into account two properties of the IN and network regions.
	First, the IN covers most of the solar photosphere at any time
	\citep[more than $90$~\% for ][which is the figure
	we use in the calculations]{Har93}. Second,
	network patches are magnetic concentrations showing a spatial 
	coherence. The network must be identified not just 
	as a strongly polarized region; it is expected to be also spatially extended.
	The procedure to define the network works as follows: (1) it automatically finds the 
	threshold on the total polarization that makes the patches of large signal to cover 
	a few percent of the FOV ($\simeq4$~\%). These signals are tentatively identified as 
    cores of network elements. (2) It 
    checks the spatial extension of the network patch candidates so as
	to exclude connected structures composed by less than ten pixels. Once
	the network patches have been identified, the procedure dilates them
	using a $10\times10$~pixel square kernel.
	The network patches thus selected cover $11$~\% of the subfield.
	
	At the location of the network patches the map showing the initial guess models
	for the inversion of each pixel reveals a significant degree of local
	coherence (Fig.~\ref{fig3}, top left panel). On the contrary, 
	the initial model MISMA fluctuates in the IN on scales
	smaller than $1''$. Considering network and IN altogether,
	the percentages of different types of  best initial guess models
	are class~0~$\simeq5\%$, 
	class~1~$\simeq18\%$, class~2~$\simeq16\%$, class~3~$\simeq10\%$,
	class~4~$\simeq11\%$, class~5~$\simeq0\%$, class~6~$\simeq7\%$,
	class~7~$\simeq12\%$, class~8~$\simeq14\%$, class~9~$\simeq5\%$, 
	and class~10~$\simeq0\%$
	(see S\'anchez Almeida \& Lites 2000 for a detailed description of the classes).
	The initial model MISMAs can be used as a proxy for the kind of asymmetries
	present in \textit{HINODE} spectra. We find that $35$~\% of the inverted profiles belong to
	classes $4$, $6$, $7$, $9$. These are the classes where Stokes
	$V$ presents large asymmetries, similar to those shown in 
	Figs.~\ref{figpm}~and~\ref{figs}. 
	
	The central panels of Fig.~\ref{fig3} show
	the magnetic field strengths
	for the two magnetized components of the 
	model MISMAs. The results have been organized so that the major and the
	minor components are considered separately\footnote{In each pixel, the major and
	the minor components are the magnetized components having the 
	largest and the smallest mass, respectively.}.
	On the one hand, the major component presents 
	occupation fractions (equivalent to volume filling factors)
	going from $\sim10^{-2}$ up to $\ga10^{-1}$,
	and magnetic field strengths between
	$0$~G and $1.8$~kG. On the other hand, the minor
	component presents filling factor almost always $\la10^{-2}$
	and field strengths $\ga1$~kG. Only in 
	the network patches the minor component presents values for the
	filling factor comparable with the values of the major component.
	
\begin{figure}[!ht]
	\centering
	\includegraphics[width=5cm]{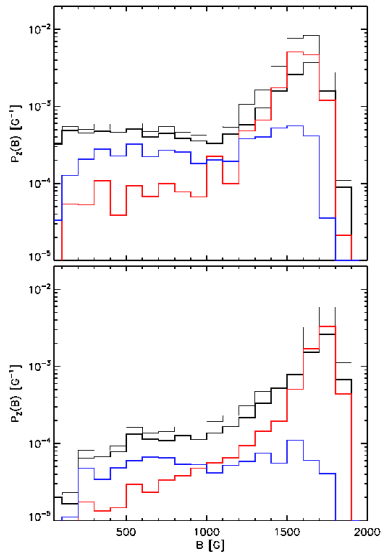}
		\caption{Fraction of analyzed photosphere covered by a given strength.
			Field strength distributions for the major component (upper panel): 
			IN (black thick line), network
			(red line), mixed polarity regions (blue line), and
        	the full $29.52''\times31.70''$ subfield (black thin line). 
        	Field strength distributions for the minor component
			(lower panel).
			}
	\label{fig4}
	\end{figure}

	The statistical properties of the field strength for both the
	major and minor components are reported in Fig.~\ref{fig4}.
    It gives the fraction of analyzed photosphere covered by
	a given field strength $P_Z(B)$, i.e., except for a scaling factor,
	it gives the sum of all occupation fractions for each magnetic
	field strength $B$.
	The histograms have been defined by
	sampling the interval $0-2$~kG with $20$ bins of $100$~G.
	They show the probability of finding a given field strength
	in our quiet Sun inversion.	
	The distribution corresponding to the whole subfield presents a
	maximum at $\simeq1.6$~kG, and then an extended tail 
	for smaller field strengths. The statistics for the
	IN (Fig.~\ref{fig4}, top, thick
	solid line) also shows kG fields, with a secondary maximum at the
	equipartition value of $\simeq5$~hG. Note the
	reduction of the kG peak with respect to the full subfield
	distribution (Fig.~\ref{fig4}, top,  thin solid line). It 
	is clear that the hG fields are almost completely in the 
	IN, while the network is dominated by kG fields with 
    little contribution for fields in the hG regime.
	On the other hand, the statistics of the minor component is
	completely dominated by strong kG fields with a maximum
	just at $1.8$~kG and almost null hG contribution; 
	this is valid for the whole subfield, IN, and
	network.
	
\begin{figure}[!ht]
	\centering
	\includegraphics[width=5cm]{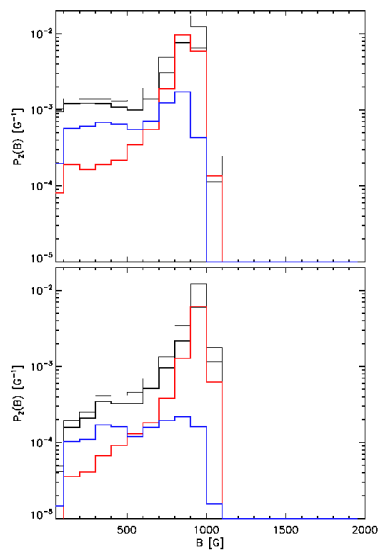}
		\caption{Fraction of analyzed photosphere covered by a given strength
			at $150$~km height in MISMA models. For details on the layout of
			the figure, see the caption of Fig.~\ref{fig4}.
			}
	\label{fig5}
	\end{figure}

	The magnetic field strength distribution in Fig.~\ref{fig5}  
	is calculated for a height in the atmosphere of approximatively
	$150$~km, 
	i.e., a height representative of the
	formation region of 
	the core of 
	the \ion{Fe}{i} lines.
	Similarly to what 
	is
	done for the definition of the reference height of
	the photosphere,
	the heights are selected as those having a total
	pressure of $4\times10^4$~dyn~cm$^{-2}$ 
	\citep[approximatively
	at 150~km in the pressure stratification of the mean
	photosphere; e.g.][]{Mal86}.
	The main difference between the two statistics in
	Figs.~\ref{fig4} and \ref{fig5} is the 
	large reduction of kG fields at $150$~km.
	The magnetic structures are treated as thin fluxtubes
	by the inversion code, therefore, the magnetic 
	field lines must fan out with height to maintain the 
	mechanical equilibrium, which implies a drop in field 
	strength, and an expansion of the magnetic
	structures \citep[e.g.,][]{spr81}.
	The two factors conspire to produce the observed
	increase of hG fields with height
	(cf. Fig.~\ref{fig4} and Fig.~\ref{fig5}).
	For example, the 
	volume occupied by magnetic fields smaller than 
	1~kG at $150$~km is some $8$ times larger than their
	volume at the base of the photosphere.

	A very interesting result, unique to MISMA inversions,
	is the detection of a large number of mixed polarity pixels 
	(three examples are reported in Fig.~\ref{figpm}).
	These are identified by the blue contours 
	on the COG magnetogram of Fig.~\ref{fig3}.
	Pixels presenting mixed polarities are very common
	in IN, namely about $25$~\% of the total number of inverted profiles.
	On the contrary, network patches tend to be interpreted by the code as unipolar regions. We 
	note that mixed polarities show up almost always either 
	in the frontiers between strong signals of opposite polarity, 
	or in low polarization regions, close
	to the $4.5\times\sigma_V$ level.
	We also note the presence of mixed polarities surrounding 
	the large negative polarity network patch in the selected subfield 
	(see the upper left corner in the COG 
	magnetogram in Fig.~\ref{fig3}). They
	are always found where positive small concentrations are close
	to the network patch.
	Fig.~\ref{fig4} also includes
	the statistical properties of the mixed polarity pixels 
	separately. Contrarily to the other statistics described above,
	the distribution for mixed polarity pixels 
	is not dominated by kG fields: hG and
	kG fields are found almost with the same 
	occupation fraction in both IN and network
	regions, and for both minor and major components.
	The distribution presents an absolute maximum at about $1.6$~kG and a 
	secondary one at about $500$~G.
	
	The fraction of analyzed photosphere occupied by magnetic fields 
	$\alpha$
	can be computed as the integral of the distributions in 
	Fig.~\ref{fig4}, i.e.,
    \begin{equation}
    \label{fill}
    \alpha=\int_0^\infty\,P_Z(B)\,dB.
    \end{equation} 
    Considering the full subfield, 
	the value of $\alpha$ for the major and the minor components 
	is $3.1$~\% and $1.4$~\%, respectively. In other words, $95.5$~\% 
	of the analyzed photosphere is field free or, more precisely, 
	it has field strengths much smaller than the ones inferred 
	from inversion, so that the inversion code cannot distinguish 
    them from zero. Other fractions corresponding to the various
    components in the FOV are listed in Table~\ref{tab}. 
    The first moment of the magnetic field strength
    distributions also provides an estimate 
    of the unsigned flux density $\langle B\rangle$
    \citep[see, e.g., ][]{DomC06b},
    \begin{equation}
    \langle B\rangle=\int_0^\infty\,B\,P_Z(B)\,dB,
    \end{equation}
    and of the average field strength,
    \begin{equation}
    \overline{B}=\langle B\rangle/\alpha.
    \end{equation}
    The values of $\langle B\rangle$ and $\overline{B}$ for the various
    components in the FOV are listed in Table~\ref{tab}. As the 
    table shows, $\langle B\rangle = 66$~G considering network and 
    IN all together. This flux density is so large partly 
    because it reflects the fraction of the $29.52''\times31.70''$ subfield
    having the 
    largest polarization signals. Such fraction $f$ 
    is about 
    $0.29$, therefore, if one considers 
    the full subfield, the fraction of photosphere occupied
    by the magnetic fields inferred from our
    inversion is $\alpha f$, and the unsigned flux 
    density $\langle B\rangle f$.  The values of the filling factor
    and unsigned flux densities corresponding to this other
    normalization are also included in 
    Table~\ref{tab} within parentheses. Then the unsigned flux density decreases 
    to about $19$~G. This still large value is put into context in 
    \S~\ref{disc}, but it is important to realize that it implicitly 
    assumes the $(1-f)$ non-inverted subfield to have no magnetic fields.
    
    We have also tried a crude separation between granules and 
    intergranules. A simple thresholding criterion was used, so 
    that pixels brighter than the mean intensity are granules, 
    and vice-versa. We find that $75$~\% of the magnetized
    plasma is in intergranules, and this fraction 
    increases with increasing field strength -- $85$~\% of the plasma having 
    $B>1.5$~kG is in intergranules. These figures are meant to be rough estimates 
    since our simple criterion is insufficient for an accurate 
    separation between granules and intergranules.
    
 \section{Discussion}
 \label{disc}
 	\begin{table*}
	\caption{
	Summary of magnetic properties at the base of the
	photosphere
	derived from the MISMA inversion\label{tab}}
	\centering
	\begin{tabular}{lcccccccccccccc}
	\hline\hline
	Portion of photosphere 	& \multicolumn{2}{c}{Field Strength $\bar{B}$~[G]} & \multicolumn{6}{c}{Flux Density $\langle B \rangle$~[G]} &
	\multicolumn{6}{c}{Filling Factor $\alpha$} \\
	& $major$ & $minor$ & \multicolumn{2}{c}{$major$} & \multicolumn{2}{c}{$minor$} & \multicolumn{2}{c}{$both$} &
	\multicolumn{2}{c}{$major$} & \multicolumn{2}{c}{$minor$} & \multicolumn{2}{c}{$both$}\\
	\hline
	internetwork (IN)		&$1258$ &$1521$	&$20$ 	&$(5.8)$	&$12$	&$(3.5)$	&$32$	&$(9.3)$	&$1.6$	&$(0.5)$	&$0.7$	&$(0.2)$	&$2.3$ 	&$(0.7)$\\
	mixed polarity regions	&$1036$ &$1041$	&$5$ 	&$(1.6)$	&$1$	&$(0.3)$	&$6$	&$(1.9)$	&$0.5$	&$(0.1)$	&$0.1$	&$(0.0)$	&$0.6$ 	&$(0.1)$\\
	network					&$1513$ &$1644$	&$23$ 	&$(6.7)$	&$11$	&$(3.2)$	&$34$	&$(9.9)$	&$1.5$	&$(0.4)$	&$0.7$	&$(0.2)$	&$2.2$ 	&$(0.6)$\\	
	full subfield			&$1380$ &$1578$	&$43$ 	&$(12.5)$	&$23$	&$(6.7)$	&$66$	&$(19.2)$	&$3.1$	&$(0.9)$	&$1.4$	&$(0.4)$	&$4.5$ 	&$(1.3)$\\
	\hline\hline
	\end{tabular}
	\note{The four rows are defined from the four distributions in Fig.~\ref{fig4}, which are normalized to the
		analyzed portion of photosphere, i.e., $29$\% of the $29.52''\times31.70''$ subfield. The values normalized to the whole subfield
		area are reported within parentheses.}
	\end{table*}

 	The ubiquity of Stokes profiles with asymmetries
 	implies that even \textit{HINODE}~SOT/SP resolution 
 	is insufficient to resolve the structure of the 
 	quiet Sun magnetic fields \citep[see][]{SanA96}. 
 	This unresolved structure has been neglected in the analyses carried out
 	so far, and we present the first attempt to account
 	for it when interpreting \textit{HINODE} spectra.
 	We employ the MISMA inversion code by \citet{SanA97},
 	which produces very good fits.
 	The MISMA description is able to reproduce the whole range of profile shapes observed 
    by \textit{HINODE}~SOT/SP in the quiet Sun, which are often extremely asymmetric. This
 	was known for Advanced Stokes Polarimeter (ASP) spectra \citep{SanALit00}, but 
 	\textit{HINODE} SOT/SP has three times the spatial resolution of
 	ASP (i.e., $0.3''$ against $1''$). 
 	
 	The percentages of MISMA classes used as initial guess
 	are different from the ones reported in \citet{SanALit00}.
 	This can be due either to the inversion strategy (e.g., the
 	imposed threshold to select invertible data) or to
 	a general modification of Stokes profile shapes due to
 	the higher spatial resolution.
 	
 	As pointed out in \citet{SanALit00}, the major component 
 	is the dominant source of polarization
 	of the model since it contains the largest mass. For this reason,
 	the statistics of the major component will be considered as the
 	reference statistics derived from our analysis.
 	The field strength distribution for the major component
 	shows an important difference with respect to the results in \citet{SanALit00};
 	in fact, our analysis is able to reveal much more hG fields than the analysis
 	of the ASP data. These are almost completely in the IN, and a large part of
 	them are found in mixed polarity pixels. 
 	Moreover, the statistics on the latter shows that hG and kG fields are 
 	measured with the same probability. The IN kG magnetic fields
 	are more important in our inversion than the inversions carried out
 	by \citet{OroS07} and \citet{AseR09}. Part of this difference is
 	certainly due to their use of ME atmospheres, unable to reproduce
 	the important line asymmetries characteristic of the quiet Sun 
 	Stokes profiles. 
 	Such difficulty may be secondary when dealing with 
 	slightly asymmetrical profiles, but the field strengths are 
 	inferred from the Stokes~$V$ shapes, and any field strength 
 	extracted by fitting anti-symmetric ME profiles to profiles 
 	like those in Figs.~\ref{figpm} and \ref{figs} is open 
 	to question.
 	Another part of the difference can come from the large 
	drop of magnetic field strength with height in the 
	atmosphere. 
	As discussed in \cite{DomC06b} and in \S~\ref{res},
	a reduction of the magnetic pressure
	is required to maintain the mechanical balance
	between the magnetic structures and the field free atmosphere
	when the gas pressure decreases exponentially with height.
	Such a decrease in field strength is intrinsic to the 
	MISMA models, which force their components to be in horizontal and
	vertical mechanical equilibrium.
	\cite{OroS07b} showed that the ME inversion of
	\ion{Fe}{i}~$630$~nm gives us information on the solar 
	atmosphere at the height of formation of the lines,
	which is higher than the reference height
	adopted in this work. When the magnetic field strength statistics
	is calculated at a height of $150$~km, then kG fields are 
	almost absent (Fig.~\ref{fig5}), which partly reconciles our 
	results with those inferred from ME inversions. However,
	the shape of our distribution still differs from
	the one reported in \citet{OroS07} in which the most probable
	field strength is $\simeq100$~G. 	
 	
 	Three additional comments regarding the field
 	strength distribution are in order.
 	First, kG fields are needed
 	to explain the presence of G-band bright points in the 
 	quiet Sun intergranular lanes 
 	\citep[e.g., ][]{SanA04,deW05,BovWie08,SanA10} and, 
    therefore, it is
    a sign of consistency that they show up in spectro-polarimetric
    observations. 
    The occupation fraction of
    kG fields inferred from the inversion 
    is of the order of a few percent, which is
	in good agreement with the $0.5-2$~\% G-band bright point 
    surface coverage  recently measured
    \citep[][]{SanA04,BovWie08,SanA10}.
    Moreover, the large polarization signals
    that we select for inversion 
    appear preferentially in the intergranular lanes 
    \citep[][]{dom03a,Lit08} 
    where the BPs are known to reside.
    Second, our results do not exclude the presence of
 	weak fields (i.e., fields producing weak polarization 
 	signals in the \ion{Fe}{i} visible lines). 
    Our inversion retrieves the field strength of
    only $4.5$~\% of the photospheric plasma (see \S~\ref{res}). 
    The field strength of the rest remains unconstrained.
    This part is almost certainly magnetic, 
    but with properties that remain elusive to our observation 
    because of its complex magnetic topology that cancels Zeeman 
    signals, because of the low field strengths, or both.	
 	\citet{SanA03} and \citet{DomC06} show how part of these hidden 
 	fields can be revealed through MISMA inversion by analyzing 
 	\ion{Fe}{i} infra-red lines. Even weaker fields 
 	can be inferred and diagnosed via Hanle effect 
    \citep[e.g., ][]{FauS94,Bom05}.
    In order to avoid overlooking the weak field component,
    the quiet Sun regions need to be analyzed
    using measurements in both visible and 
 	infra-red spectral ranges, and produced by both Zeeman
 	and Hanle effects. 
    How to consistently combine these measurements remains 
    an open issue, but such combination is certainly the way to go 
    \citep[see the first attempt by ][]{DomC06b}.
    The third comment on the field strength distribution refers to
    discarding biases that may artificially produce kG fields. 
    \citet{BelRColl03} show how large 
    noise in the Stokes $V$ profiles of the \ion{Fe}{i}~$630$~nm lines 
    may fake kG fields. However, this effect is insignificant above 
    a signal-to-noise ratio threshold that our typical
    spectra exceed. Moreover, in addition to kG, noise also produces
    false weak dG, which we do not see. Another potential source of 
    false kG fields could be (polarized) stray-light from the 
    network to the IN. Although we cannot discard some influence 
    close to the network patches, stray-light cannot account for the 
    bulk of the kG fields in the IN. Stray-light is at most 10\% as 
    measured for the broadband filter imager aboard \textit{HINODE} 
    \citep[][]{WedB08}. 
    It could account for 10\% of the network magnetic flux 
    artificially appearing in the IN, but this represents only a 
    few~G, thus unable to explain the some 30~G inferred for the IN 
    (see Table~\ref{tab}).
 	
 	The mixed polarity regions revealed by  the MISMA
 	analysis are about $25$~\% of the total number of inverted profiles,
 	similar to the percentage reported in \citet{SanALit00} 
 	for ASP data \citep[see also ][]{SocNSanA02}.
 	Note that the percentage of mixed polarities has not
 	changed despite the increase of angular resolution (from
 	$1''$ to $0.3''$), and despite the fact that 
 	the polarization signals analyzed here are generally
 	larger (we use a noise threshold $1.5$ times larger
 	than the one used on ASP spectra). The really significant
 	difference between the two datasets is the fraction 
 	of solar surface covered by polarization signals; 
  	\citet{SanALit00} invert only $15$~\% of the surface, 
  	whereas \textit{HINODE} spectra allow us to almost double this 
  	fraction. The increase of angular resolution has certainly
  	resolved some of the mixed polarities detected at ASP
  	angular resolution, however, at the same time, the 
  	newly revealed Stokes $V$ profiles uncover mixed 
  	polarities at even finer spatial scales. 
  	This observation is consistent with the numerical models of solar
  	magneto-convection and/or turbulent dynamos 
  	\citep[e.g., ][]{Cat99,VogSch07}. They predict an intricate
 	pattern of highly intermittent fields varying over very small
 	spatial scales down to the diffusion length-scales.
 
 	The average flux density we infer from the inverted profiles
 	is of the order of $19$~G when the area of the full subfield
 	is considered for normalization (see Table~\ref{tab}). 
 	This value is almost twice as large as the one inferred by 
 	\citet[][$1''$ resolution]{SanALit00}, and it is also
    significantly larger than the ones 
    obtained from ME inversions of quiet Sun \textit{HINODE} spectra
    by \citet[][$\sim9.5$~G for the full FOV]{OroS07}, 
    and using the magnetograph equation by 
    \citet[][$11$~G]{Lit08}.
    However, our estimate is comparable with other \textit{HINODE} 
    based estimates of the unsigned flux of the quiet Sun
    (e.g., Jin et al. 2009 get $28$~G).	
	Such a difference could be explained taking into account 
	that ME inversions are blind to sub-pixel magnetic structuring. 
	Note that the average flux density from the major component 
	(i.e., the dominant source of polarization) is in good 
	agreement with \citet{OroS07} and \citet{Lit08} --
	we get $12.5$~G, see Table~\ref{tab}. The small excess with respect
	to their results is compatible with the $1.6$~G flux contribution
	from mixed polarities (see Table~\ref{tab}). The bulk of the
	difference seems to reside in the $6.7$~G provided by the minor 
	component. Its contribution to the flux density is
	more important than the modification it produces on the 
	polarization signals (even though it is needed to reproduce the 
	line asymmetries). 
 	
\section{Conclusions}
\label{conc}
	We present the first inversion of \textit{HINODE}~SOT/SP 
	Stokes profiles that accounts for the asymmetries of the 
	profiles in the quiet Sun.
	The analysis is carried out under the MISMA hypotheses
	which allows to reproduce the different profile shapes observed in
	the quiet Sun.
	We follow the approach  already successfully exploited to describe the 
	asymmetries observed with $1''$ angular resolution 
	\citep{SanALit00}. 
	
	The main results are as follows:
	\begin{enumerate}
	\item We inverted 11600 sets of Stokes $I$ and $V$ profiles 
	representative of the quiet Sun internetwork (IN) and network at disk center.
	The inversion code is able to reproduce, in a satisfactory way, the whole
	variety of asymmetries revealed by \textit{HINODE}~SOT/SP.
	
	\item The MISMA code is also
	able to reproduce linear polarization measurements
	when full Stokes inversions are performed.
	Here we report a few examples.
	 
	\item The existence of asymmetries is certainly not 
	negligible. Some $35$~\% of the analyzed profiles
	present large asymmetries, according to the 
	rough classification used in \citet{SanALit00}.

	\item $25$\% of the analyzed profiles present
	asymmetries that are interpreted by the MISMA code
	as due to mixed polarities in the resolution element.
	These pixels are found to be located either in 
	transition regions between patches of opposite polarity,
	or in pixels presenting weak polarization signals.

    \item The magnetic plasma whose properties the inversion
    code constrains represents only $4.5$~\% of the photospheric
    plasma. The rest remains unconstrained by our analysis. 

	\item
	The statistics of the detected magnetic field 
	strength at the formation height of the continuum
	is dominated by strong kG 
	fields, both in the network and the IN. The 
	later, however, presents an extended tail
	over the whole hG domain.
	
	\item 
	At the height of 
	$150$~km (representative of the formation
	heights of the \ion{Fe}{i} visible lines) very little 
	kG fields are measured.
	This result 
	narrows down the gap between our field strength distribution
	and those inferred from ME inversions by
	\citet{OroS07} and \citet{AseR09}, and it
	follows directly from the decrease of the
	field strength in response to the decrease of the gas pressure 
	with height 
	in the photosphere.
	
    \item The average flux density derived from the
    inverted pixels is $66$~G. If one considers the 
    full analyzed subfield, it becomes $19$~G. The same figures
    for the IN become $32$~G and $9.3$~G, respectively. 
    These values are significantly larger than the ones 
    obtained from ME inversions of quiet Sun \textit{HINODE} spectra
    by \citet[][$\sim9.5$~G for the full FOV]{OroS07}, 
    and using the magnetograph equation by 
    \citet[][$11$~G]{Lit08}.
    However, our estimate is comparable with other \textit{HINODE} 
    based estimates of the unsigned flux of the quiet Sun 
    (e.g., Jin et al. 2009 get $28$~G).
	\end{enumerate}
	
	As a general 
	concluding
	remark, the analysis here
	presented reveals the importance of a complete interpretation of
	the shape of Stokes $V$ profiles through 
	refined inversion codes. 
	This work should be considered
	as the first step in the interpretation of \textit{HINODE}
	SOT/SP profile shapes. 
	Because of the wealth of information contained in the asymmetries,
	such a topic deserves attention from the solar community.
	Different inversion codes and hypotheses
	can be adopted to interpret such asymmetries, and they
	will probably be able to reproduce the line shapes as well
	\citep[e.g., a systematic variation 
	of magnetic field and velocity along the LOS;][]{rui92}.
	The uniqueness of the interpretation cannot be assessed
	unless different alternatives agree.	
	
\acknowledgements
	\textit{HINODE} is a Japanese mission developed and launched by ISAS/JAXA, 
	with NAOJ as domestic partner and NASA and STFC (UK) as 
	international partners. It is operated by these agencies 
	in co-operation with ESA and NSC (Norway).
	JSA acknowledges the support provided by the Spanish Ministry 
    of Science and Technology through project AYA2007-66502, as well as 
    by the EC SOLAIRE Network (MTRN-CT-2006-035484). This work
    was partially supported by ASI grant n.I/015/07/0ESS.


\appendix

\section{Lack of crosstalk between magnetic field strength and magnetic
field inclination}
\label{appendix2}
\begin{figure}
	\centering
	\includegraphics[width=5cm]{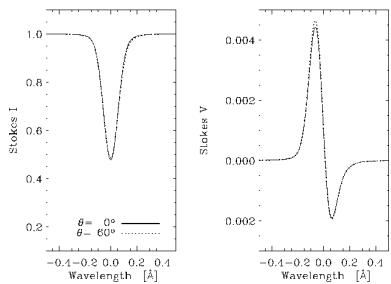}
	\caption{
		Two sets of synthetic Stokes $I$ and $V$ profiles of 
		\ion{Fe}{i}~$630.25$~nm 
		obtained with different magnetic field inclinations $\theta$
		but the same product $(1-f_{sl})\,\cos\theta$, with $f_{sl}$ the fraction of 
		stray-light. The solid and the dotted lines correspond to $\theta=0$\degr and $60$\degr, 
		respectively.
		The two sets cannot be distinguished within typical \textit{HINODE} noise. 
		}
	\label{fig_app}
	\end{figure}

	When $Q$ and $U$ are too small to be used,  
	the inversions assume the magnetic field to be longitudinal.
	It is important to realize that such assumption does not 
	influence our magnetic field strength diagnostics. 
	When the polarizations signals are weak (our case), and
	the inclination is constant (our assumption), then Stokes~$V$ scales
	with the cosine of the inclination independently of the field
	strength \citep[][\S~3.1]{SanATruB99}.
	The information on the field strength is coded in the shape of 
	Stokes~$V$, whereas the magnetic field inclination modifies
	the scaling. Such scaling 
	is
	independently obtained
	by the inversion code disguised as stray-light, therefore the 
	inferred magnetic field strength and inclination are uncoupled. 
	All quiet Sun inversions assume the pixels to 
	be partly covered by magnetic fields, and the corresponding
	filling factor is determined as a free parameter of the
	inversion (stray-light factor). An error on the
	magnetic field inclination affects the stray-light factor, 
	but it does not influence the magnetic field determination. 
	Figure~\ref{fig_app} illustrate the argument. It shows Stokes $I$
	and $V$ synthesized in two atmospheres that differ
	in stray-light factor $f_{sl}$ and cosine of inclination
	$\cos\theta$, but they have the same product $\cos\theta\,(1-f_{sl})$.
	The two pairs of Stokes profiles are indistinguishable
	within the noise of \textit{HINODE} observations.
	
\end{document}